\documentclass[prodmode]{acmsmall} 

\usepackage[ruled]{algorithm2e}

\SetAlFnt{\small}
\SetAlCapFnt{\small}
\SetAlCapNameFnt{\small}
\SetAlCapHSkip{0pt}
\IncMargin{-\parindent}

\usepackage{subfigure}

\acmVolume{9}
\acmNumber{4}
\acmArticle{39}
\acmYear{2010}
\acmMonth{3}

\doi{0000001.0000001}

\issn{1234-56789}

\begin{document}

\markboth{Singh et al.}{Learning User Representations using Temporal Dynamics of Information Diffusion}

\title{Learning User Representations in Online Social Networks using Temporal Dynamics of Information Diffusion}
\author{HARVINEET SINGH
\affil{Indian Institute of Technology, Delhi}
AMITABHA BAGCHI
\affil{Indian Institute of Technology, Delhi}
PARAG SINGLA
\affil{Indian Institute of Technology, Delhi}}

\begin{abstract}
This article presents a novel approach for learning low-dimensional distributed
representations of users in online social networks.
Existing methods rely on the network structure formed by the social relationships
among users to extract these representations.
However, the network information can be obsolete, incomplete or
dynamically changing. In addition, in some cases it can be prohibitively expensive 
to get the network information. Therefore, we propose an alternative approach
based on observations from topics being talked on in social networks. 
We utilise the time information of users adopting topics in order to embed 
them in real-valued vector space.
Through extensive experiments we investigate the properties of the
representations learned and their efficacy in preserving information about
link structure among users. We also evaluate the representations in two
different prediction tasks, namely, predicting most likely future adopters
of a topic and predicting the geo-location of users.
Experiments to validate the proposed methods are performed on a
large-scale social network extracted from Twitter, consisting of about
7.7 million users and their activity on around 3.6 million topics over a month long period.
\end{abstract}

%
%


%
%



\acmformat{Harvineet Singh, Amitabha Bagchi,
and Parag Singla, 2015. Learning Representations of Users in Online Social Networks using Temporal Dynamics of Information Diffusion.}

\begin{bottomstuff}
Author's addresses: Harvineet Singh, Amitabha Bagchi  {and} Parag Singla,
Indian Institute of Technology, Hauz Khas, New Delhi, India.
\end{bottomstuff}

\maketitle

\section{Introduction}
\label{sec:intro}

Predicting the future behaviour or attributes of users of online social networking services such as \textit {Facebook}, \textit {Twitter}, \textit {LinkedIn} etc. has been a focus of many recent studies. Examples include work on link prediction, modelling of spread of information, content recommendation, predicting user demographics among others. Increased observability of the microscopic interactions in these online platforms supports richer analysis of these prediction tasks.\\
One important pre-processing step employed in such prediction tasks is feature extraction. The aim here is to represent the raw input data in such a way that it can be effectively leveraged by the learning models. Traditionally, this involves hand crafting a set of features that capture relevant discriminatory information from the observed data. There are many potential shortcomings of this approach. In addition to involving excessive trial-and-error, this approach also relies on extensive domain knowledge to filter through irrelevant information in the observed data.
Also, the set of features extracted for a particular task may not be generalise to other tasks of interest. Thus, leading to duplication of effort.\\
An alternate approach, which aims at overcoming some of these problems, is to \textit{learn good representations} of the raw data by defining supervised or unsupervised tasks over the data. Thus the process of identifying features is automated by making it a part of the learning algorithm itself instead of hand-crafting them. \textit{Representation learning} involves a set of methods that learn representations of the raw data that can be effectively leveraged by standard machine learning algorithms~\cite{bengio2013representation}.
An example of such a method is Principal Component Analysis (PCA) which learns representations of the input data points in a lower dimensional space that explain maximum amount of variability observed in the input. Another popular approach employs multilayer neural networks to learn representations or features from raw data instead of hand-crafting them. This has been successfully used in recent works in computer vision, speech recognition and natural language processing \cite{lecun2015deep}.\\
We explore the use of such methods to learn representations of nodes in social networks. The data from social networks is characterised by presence of different types of entities (like users, content), different attributes associated with them (like demography of users, their friendship information, topics discussed). For a given task, taking into account the complex interactions present among these entities in order to extract features manually can be a difficult task. We hypothesise that the representation learning approach can capture these complex interactions and leverage the rich information present in social networks which can be difficult to capture otherwise or may involve a lot of trial and error to extract an informative set of features for a given task.\\
Our methodology is based on a technique from statistical language modelling, referred to as \textit{word2vec}, which is used for learning vector representations of words from input text corpus. This has been successfully used in many diverse applications such as for natural language processing tasks like machine translation \cite{mikolov2013exploiting}, product recommendations \cite{grbovic2015kddB},  social network analysis \cite{perozzi2014deepwalk} among others.

\paragraph{Organization} We discuss the related work in
Section~\ref{sec:related}, following that with a description of our
dataset in Section~\ref{sec:dataset}. Our methodology for
extracting user representations is described in Section~\ref{sec:method},
followed by a detailed discussion of their characteristics
in Section~\ref{sec:eval}. The task definition and results
of evaluation of learned representations on two prediction tasks
is presented in Section~\ref{sec:pred}. Finally we conclude with a 
discussion of the findings and some future directions in
Section~\ref{sec:conclusion}.

\section{Related work}
\label{sec:related}

There has been much recent interest in learning node representations in social networks. These representations are \textit{learned} to capture the observed interactions among nodes, e.g. their link structure. \cite{tang2009relational} extract top-\textit{k} eigenvectors of the modularity matrix of the network graph to get latent features for users and utilise them as features for classification tasks. Another work introduces the existing representation learning methods used in natural language processing for learning representations in social networks \cite{perozzi2014deepwalk}. They use multiple short-truncated random walks starting from nodes in the network to generate sequences of nodes. These can be considered analogous to sentences in text. Then, \textit{Skip-gram model} \cite{mikolov2013distributed} is used on the resulting corpus of sentences to represent each node as a real-valued vector. The learned representations capture meaningful information about the network structure as demonstrated using node classification tasks. \cite{tang2015line} propose a scalable embedding method with an objective function that explicitly encodes the \textit{first-order} and \textit{second-order proximity} information of users. The first-order proximity between a pair of nodes refers to information on their direct connectivity (i.e. whether they are connected or not and edge-weights in case of weighted networks), whereas second-order proximity refers to similarity in first-order proximity between nodes. An optimisation criteria is proposed that preserves both kinds of information of the nodes, while representing them in a low-dimensional space. \cite{grarep2015} propose a method to learn graph representations that allows to take into account higher-order proximities.\\
However, these methods assume the knowledge of the network structure which might not always be available or even observable. Moreover, the link structure might not give a true picture of the interactions among users \cite{huberman2008social}. Also, it is highly dynamic. The structure of the network can change as users add and/or delete edges in response to various factors including exposure to content shared by others \cite{myers2014bursty}. On the other hand, the timing information of topic adoptions is readily available and can be used instead of relying on the link structure, e.g. \cite{gomez2012inferring} used timing information of node adoptions to infer the diffusion network, i.e., who influenced whom to adopt the information.\\
The present work uses the node adoption times to learn real-valued vector representations of the participating nodes. In a closely related work, \cite{bourigault2014learning} propose an approach that does not depend on the knowledge of the network, but instead uses the timestamps of adoption from the observed cascades (sequence of users adopting a topic). The diffusion of information is modelled using a \textit{heat diffusion kernel}.
The aim is to learn the vector representations of users, namely parameters of the kernel, that best explains the temporal ordering of the observed cascades. The learned parameters are then used to predict future adopters of a cascade given its source user. In contrast, our approach first uses the time information of topic adoptions to construct a graph and then employs techniques from statistical language modelling to learn representations using this graph.
\section{Dataset Description}
\label{sec:dataset}

Our dataset comprises of the complete set of tweets posted by around 7.7 million Twitter users in a month long period between 27 March, 2014 and 29 April, 2014. We also obtained the followee-follower relationships among these set of users using Twitter REST API. In total, the dataset consists of about 0.2 billion tweets containing about 8 million hashtags. Along with the text of the tweets, the time zone associated with the twitter account of each user, the type of tweets posted (\textit {replies}, \textit {mentions}, \textit {retweets}) and the time of posting were also extracted. This allowed us to construct a detailed view of user activity on each topic in the dataset for a period of one month. The procedure adopted for data collection is detailed in \cite{bora2015role}. Table~\ref{tbl:dataset} contains some basic statistics of our dataset.

\begin{table}[htbp]
\small
\centering
\tbl{Dataset statistics\label{tbl:dataset}}{
\begin{tabular}{|l|r|}
\hline
Users & 7,695,882\\
\hline
Average no. of followers & 450 \\
\hline
Users who tweet at least once & 3,008,496 \\
\hline
Number of hashtags & 8,793,155\\
\hline
Number of tweets & 220,012,557 \\
\hline
\end{tabular}}
\end{table}

\section{Learning User Representations}
\label{sec:method}
We used Skip-gram model to learn user representations, as proposed in \cite{perozzi2014deepwalk}. In contrast to using link structure, we utilised topic adoption sequences which provide richer user interaction information and allowed us to model information diffusion. We discuss the Skip-gram model and then describe our methodology in more detail.

\subsection{Skip-gram model}
Skip-gram model proposed in (\cite{mikolov2013efficient},\cite{mikolov2013distributed}) is a neural network based model for learning \textit{distributed} vector representations of words. Here, distributed means that there exists a many-to-many mapping between the dimensions of the vectors and the properties of words which the representation tries to encode. This is different from the traditional approach of representing words as \textit{1-of-N} vectors which encodes a word's position in the vocabulary as 1 and rest of the values are 0. Thus, this encoding scheme doesn't explicitly capture the similarity between words.\\
Skip-gram model defines an objective function that trains the word vectors to predict their context, where context is defined as the words occurring within some fixed distance of the given word in the training corpus. Given a word $u$ and a word $c$ in its context, the conditional probability $p(c|u)$ is given by the \textit{softmax function},
\begin{equation}
p(c|u)=\frac{exp\ (v'_c.v_u)}{\sum\limits_{w\in V}exp\ (v'_w.v_u)}
\end{equation}
where $v_w$ and $v'_w$ are vector representations of word and context respectively which are  parameters of the model, $V$ is the set of words in the training corpus.\\
The training objective is to maximise the sum of log probabilities of all word-context pairs in the corpus:
\begin{equation}
\sum\limits_{(u,v)\in C}log\ p(v|u)
\end{equation}
where $C$ is the set of all word-context pairs.\\
\cite{mikolov2013efficient},\cite{mikolov2013distributed} describe various techniques for efficient optimisation of this objective. The method is highly-scalable enabling word vectors to be learned from large datasets containing billions of words. As a result of the training objective chosen, words occurring in similar context are represented by similar vectors. The word vectors thus obtained have been shown to capture semantic and syntactic relationships among words.

\subsection{Methodology}
We used hashtags tweeted by the users to extract context for them. Users tweeting on the same hashtags indicate a measure of similarity among them. This similarity can be exploited for embedding users. We first describe the procedure for extracting contexts from users' tweets on hashtags.\\
For a hashtag $h$, the adoption sequence $S_h$ is the time-ordered sequence of users' tweets on $h$. For each tweet in $S_h$ and the corresponding user $u$, the context for $u$ is given by the users in $S_h$ who tweeted within a time period $\tau$ from $t(u)$. Here, $t(u)$ is the timestamp of the tweet by $u$. If a user tweets multiple times on a hashtag, multiple such contexts are extracted. Repeating this for all hashtags gives user-context pairs for all users. This definition of context tries to capture the users who have similar adoption patterns, i.e. similar in the topics they tweet on and the time at which they tweet.\\
If the tweets on a hashtag are tweeted very close by in time, this will result in a large number of contexts. One way of limiting the contexts is by varying $\tau$ for each tweet. We instead used a \textit{path sampling} procedure. This converts $S_h$ into a directed graph $G_h$, where the nodes are the tweets of $S_h$ and an edge exists from $a$ to $b$ if $t(b)-t(a) \le \tau$. Starting from a node in $G_h$, the next node in the path is uniformly sampled from its neighbours. This is done repeatedly until path of length $\gamma$ is obtained (this is same as using a fixed-length random walk). All the nodes in the path are considered as contexts for each other. A single path is sampled for each node in $G_h$. This sampling procedure limits the number of contexts for a user. This also has the advantage of capturing users tweeting at time differences larger than $\tau$ as context. Skip-gram model is then used to learn $d$-dimensional user vector representations from the extracted contexts.\\ 
\paragraph{Training details} For training, \textit{word2vec} toolkit\footnotemark \footnotetext{https://code.google.com/p/word2vec/} was used. The time difference for an edge $\tau$ was set to 1 hour and path length $\gamma$ was set to 10. Parameters for word2vec used were: vector dimension $d$ 100, Skip-gram with Hierarchical Softmax, context window length 10, sub-sampling threshold $10^{-4}$, training iterations 20, default values for rest of the parameters were used. The training took approx. 1 day on a 128 GB machine using 10 cores.
\section{Properties of representations}
\label{sec:eval}

We now look at the properties of users that are preserved by these embeddings. In order to do this, we compare the neighbours of users in the vector space with their neighbours in the followee-follower network. The network neighbourhood exhibits high degree of similarity among users as the links are formed on the basis of friendships, shared interests, etc. (a phenomenon referred to as \textit{homophily} i.e., tendency of users to form links with similar users). On the other hand, the embeddings are obtained solely on the basis of topic adoptions without considering the network structure. Therefore, we test for the similarity of the two different kinds of neighbourhoods.\\
In particular, we look at the following questions,
\begin{enumerate}
\item How similar are these neighbourhoods w.r.t. the users contained in them?\\
For a particular user, we first obtain the set of its followers and the set of its k-nearest neighbours (where, k is taken to be same as number of followers) in the vector space. These two sets are then compared using Jaccard similarity index (given as cardinality of intersection of the sets divided by cardinality of union of the two). This gave a low similarity index of about $0.01$ (averaged over $1000$ randomly sampled users). Thus, in terms of the users contained in them, the two neighbourhoods are different.
\item For a user, whether its network neighbours are more likely to co-adopt topics than its neighbours in vector space?\\
Firstly we define the likelihood of co-adoption for a user and its neighbourhood as follows: total number of times the user and any of its neighbours adopted the same topic divided by the total possible number of such co-adoptions, which is, number of neighbours multiplied by number of topics adopted by the user.
\begin{equation}
p_u=\frac{\sum\limits_{w\in N(u)} \sum\limits_{t\in T(u)} \textit{I}\ (w\ adopts\ t)}{|N(u)|.|T(u)|}
\end{equation}
where, $p_u$ is the likelihood of co-adoption of user $u$, $N(u)$ is the set of its neighbours, $T(u)$ is the set of topics adopted by $u$ and \textit{I} is the indicator function.\\
We compute this measure for both network and vector space neighbours for a random sample of $10000$ users. On comparison, the vector space neighbours had higher likelihood of co-adoption than network neighbours on average ($0.053$ and $0.0359$ respectively). The same can also be observed in Figure~\ref{fig:coadopt} from the skew of the points.
\begin{figure}[ht!]
\centering
\includegraphics[width=.65\linewidth]{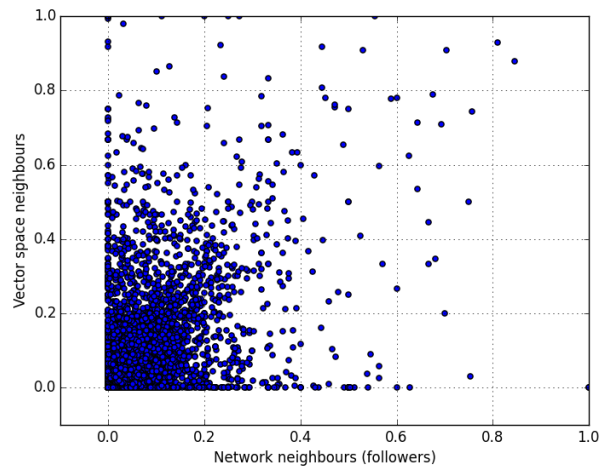}
\caption[Likelihood of Co-adoption]{Scatter plot of the likelihood of co-adoption of users with different neighbourhoods. The distribution is skewed towards the upper side of the $1$:$1$ line.}
\label{fig:coadopt}
\end{figure}
This also gives us an intuition into how the embeddings can be used in the adopter prediction task. Instead of looking at the network neighbours of the initial adopters of a topic in order to predict its future adopters, we can work with their vector space neighbours, as these have higher likelihood of co-adoption.
\end{enumerate}

\begin{figure*}[ht!]
\centering
\begin{subfigure}[]{
\includegraphics[width=.65\textwidth]{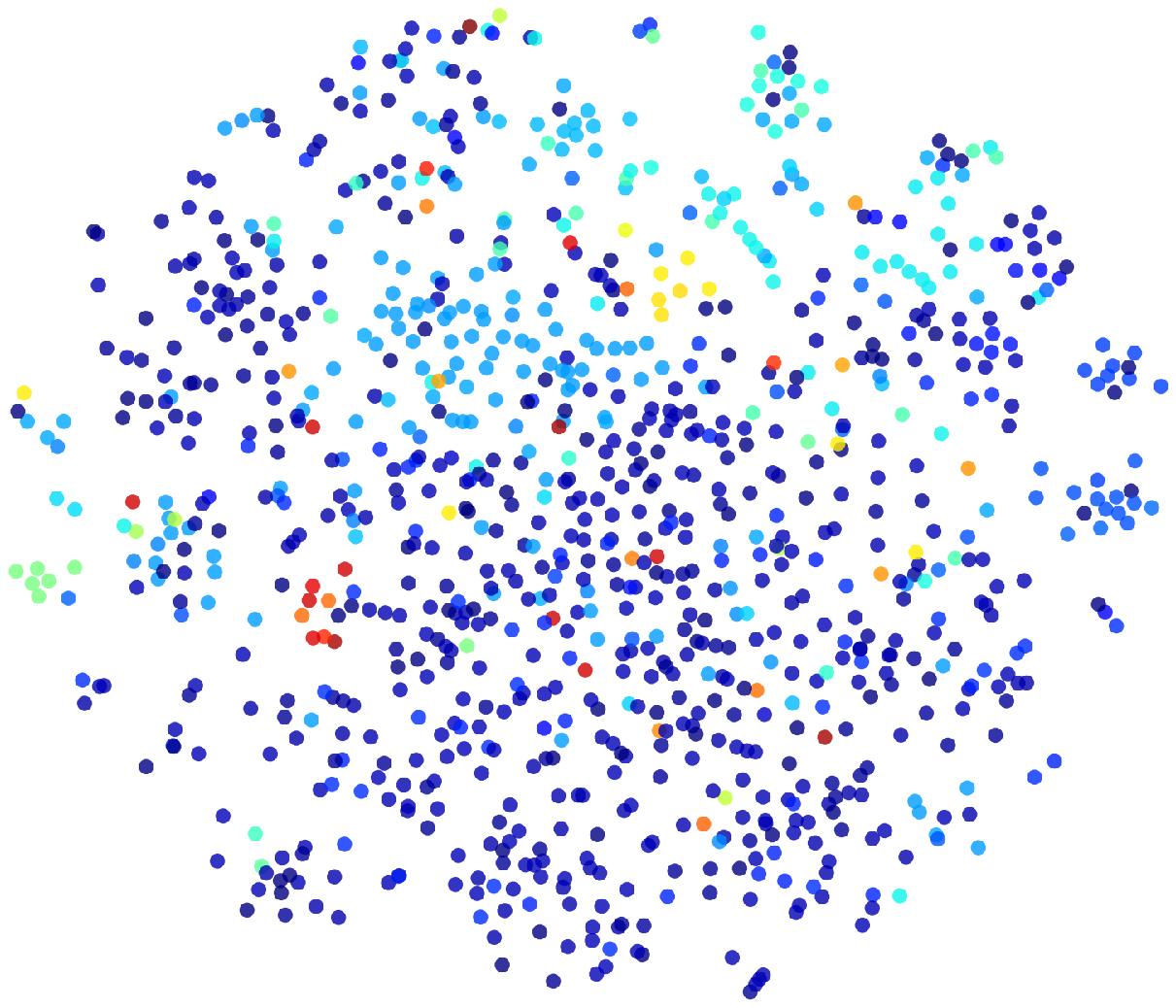}
\label{fig:geovis}
}
\end{subfigure}
\begin{subfigure}[]{
\includegraphics[width=.45\textwidth]{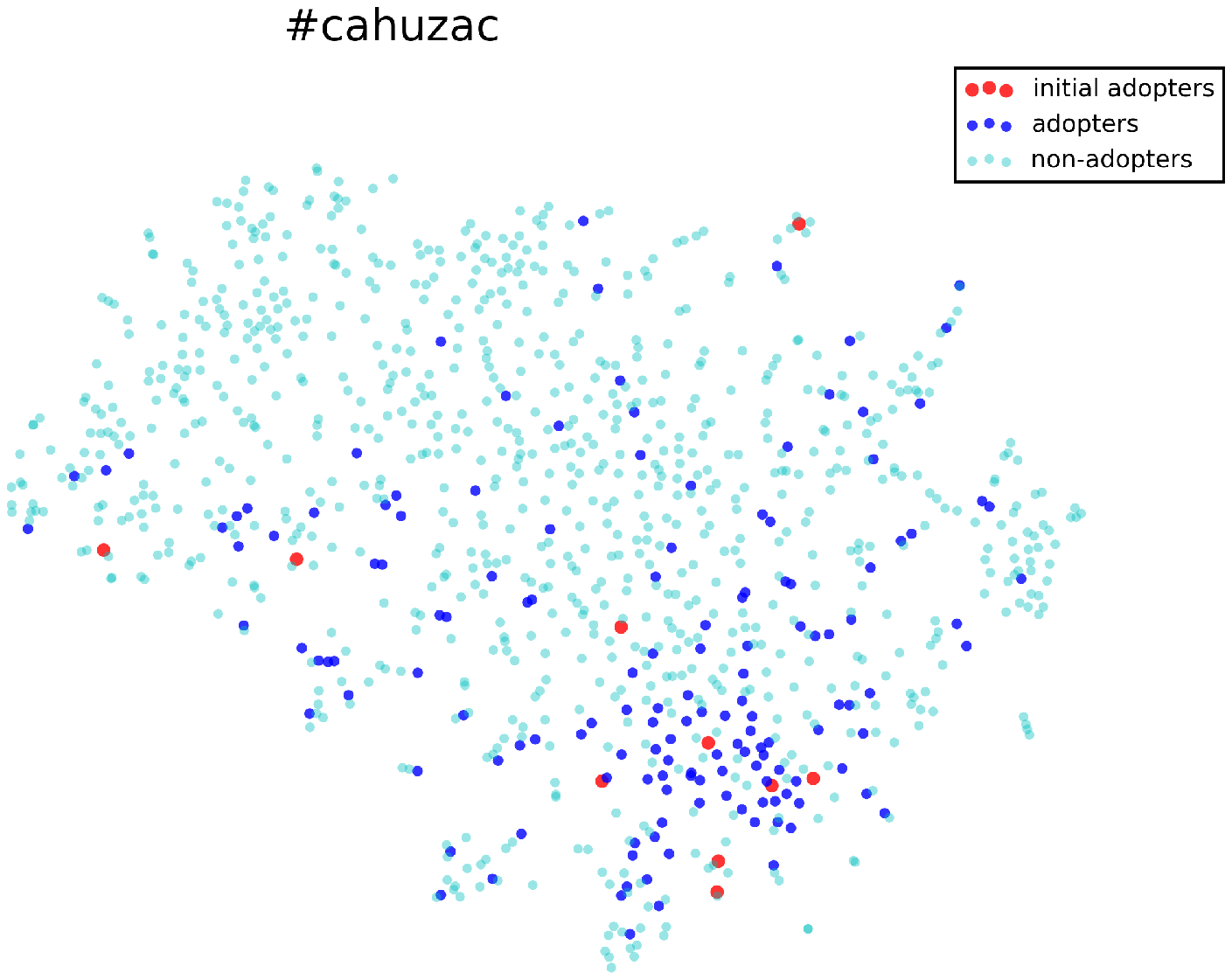}
\label{fig:vis1}
}
\end{subfigure}
\begin{subfigure}[]{
\includegraphics[width=.45\textwidth]{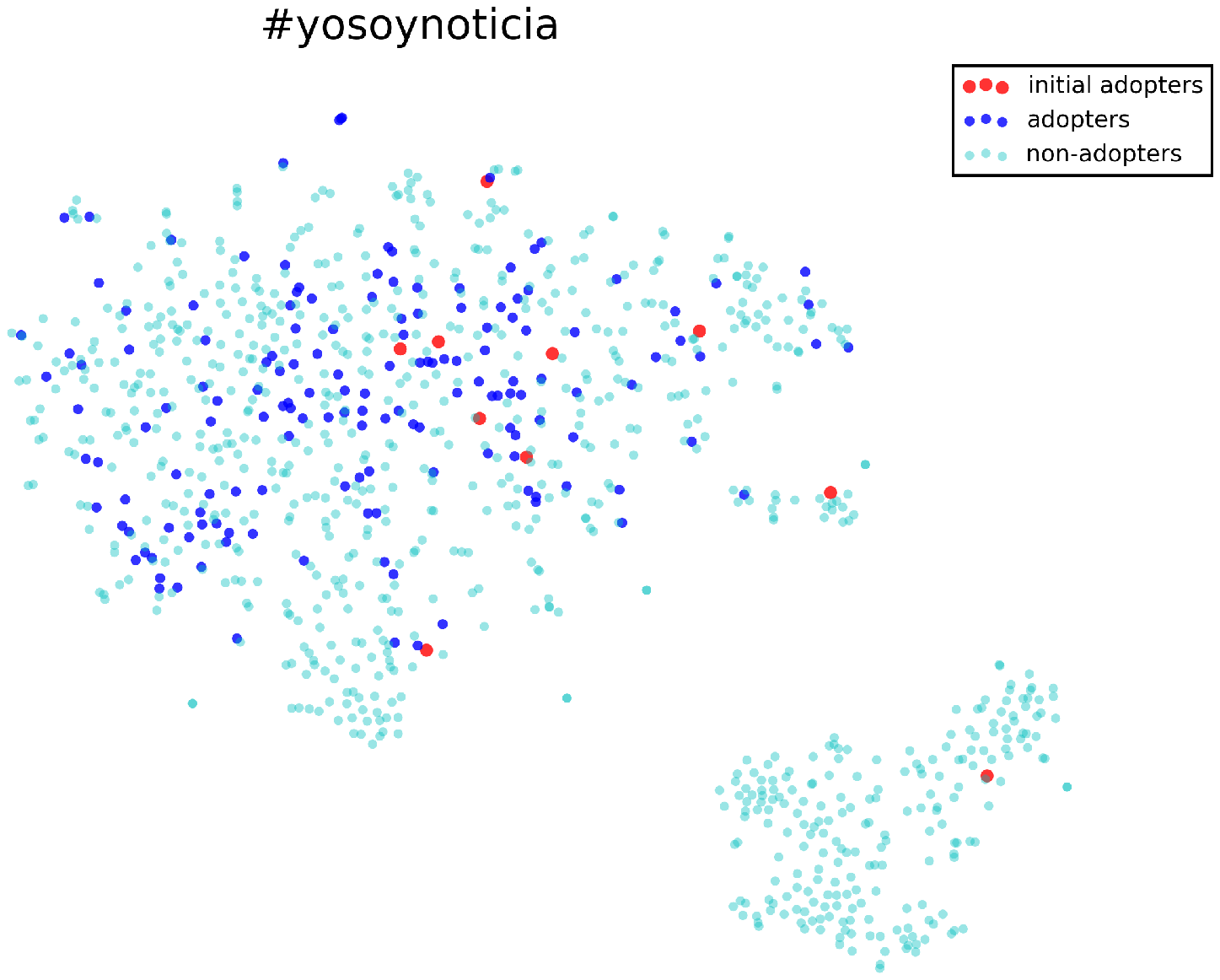}
\label{fig:vis2}
}
\end{subfigure}
\caption{(a) Visualisation of embeddings where colour represents the geographical location of the user. 1000 randomly sampled users are plotted. (b), (c) Visualisation of the neighbourhood of initial adopters of a topic.}
\label{fig:vis}
\end{figure*}

In order to visualise the user embeddings, we used a dimensionality reduction technique called \textit{t-SNE} \cite{van2008visualizing} to get a two-dimensional representation, as shown in Figure~\ref{fig:geovis}. Since, the embeddings are obtained using the spread of topics, we visualise this spread by plotting the embeddings of the topic's adopters. To contrast the adopters with non-adopters, the first $10$ adopters of the topic are taken and $1000$ nearest neighbours of these users are queried using their vector space representations. Adopters and non-adopters from this neighbourhood have been identified separately in Figure~\ref{fig:vis1} and \ref{fig:vis2} for two sample topics from the dataset.
 
\section{Evaluation of representations}
\label{sec:pred}

To evaluate the versatility of learned representations, we test them on two prediction task, namely,   predicting geo-location of users and predicting future adopters of a topic.

\subsection{Geo-location Inference}
The time zone information of the Twitter profile of the user was taken as its geo-location. This is relatively coarse level of geographical information. While geo-tagged tweets enable finer geographical information, these were not available for most of the users. The prediction task was formulated as that of multi-class classification (number of time zones were 141). User representations were used as features for classification and the time zone of each user was taken as the target variable. 

\subsubsection{Experiment}
We used \textit{one-vs-rest} logistic regression model implemented in LibLinear toolkit \cite{fan2008liblinear} for classification. Users were randomly split into train and test sets with varying sizes of the training set. Overall accuracy on the test set (ratio of correctly predicted examples to the number of examples) was taken as the evaluation metric. Results were computed on a random sample of $1$ million users with known geography.\\
For comparison, a network-based baseline was considered. This method infers the geography of a user from the most frequent geography among its neighbours \cite{davis2011inferring}. The neighbours are considered as the users followed by the given user, i.e. its "friends" in Twitter terms. Ties are broken arbitrarily. If none of the neighbours' location is known for a user, then the most-frequent location in the training set is predicted. Majority guess baseline predicts the most frequently seen location in the training set for all users.

\begin{table}[ht!]
\centering
\tbl{Geo-location inference results\label{tab:geopred-user}}{
\begin{tabular}{|l|l|l|l|}
\hline
\begin{tabular}[c]{@{}l@{}}{\bf Model/\% of training data}\\ {\bf (Accuracy \%)}\end{tabular} & {\bf 1\%}   & {\bf 5\%}   & {\bf 10\%}  \\ \hline
Majority guess baseline                                                           & 19.64 & 19.67 & 19.61 \\ \hline
\begin{tabular}[c]{@{}l@{}}Network-based baseline\\ (Friends)\end{tabular}        & 25.40 & 36.58 & 40.52 \\ \hline
\begin{tabular}[c]{@{}l@{}}User vectors\end{tabular}      & 38.76 & 40.30 & 40.58 \\ \hline
\end{tabular}}
\end{table}

\subsubsection{Results}
Table~\ref{tab:geopred-user} details the results of the experiment when the size of training set used is varied from 1\% to 10\%. The difference in performance is not significant when location of high percentage of users is known. However, user vector based method out-performs the baselines when less information is available. Even though the representations are not trained for this task explicitly but the good prediction performance indicates the generality of these representations. The classifier is able to generalise well from limited information of the class labels of points in the feature space. This could be because the structure of feature space is such that the users with same geography are embedded near each other.\\
While the tweets and their timestamps are readily available, the geo-location information may not be. Thus, the learned features can be used to predict geography of the users from the geographical information of only a small number of users.

\subsection{Adopter Prediction Task}
Given the initial set of adopters of a topic, the task is to predict it's subsequent adopters.

\subsubsection{Methodology}
Given the first $n$ adopters of a topic, the model predicts $k$ users most likely to adopt it in future. The task can be seen as that of information retrieval, where the \textit{query} consists of the set of first $n$ adopters $S$ and the \textit{relevant documents} to be retrieved are the future adopters of the topic. For this retrieval, we need a measure to rank candidates according to their relevance to the query. We use the learned vector representations for ranking candidates. Here, candidates are taken as all the users except the initial adopters.\\
The ranking criteria uses a score assigned to each candidate user $c$ which is based on the Euclidean distance of $c$ from each user in $S$, measured using their corresponding vectors. We consider different ways of combining these distances:
\begin{itemize}
\item \textbf{Min:} $score(c,S)=\min\limits_{a \in S} d(c,a)$,  i.e. candidates are ranked in increasing order of their minimum distance to $S$. An implementation detail: For getting the top-\textit{k} ranked list according to this criteria, \textit{k}-nearest neighbours were queried for each user in $S$ using \textit{k}-d tree for efficient indexing and these lists were then merged. This provided a speed-up of about 10x compared to brute-force search, despite the high-dimensional data points.
\item \textbf{Average:} $score(c,S)=\frac{1}{|S|} \sum\limits_{a \in S} d(c,a)$, i.e. candidates are ranked in increasing order of their average distance to $S$. This criteria involves computing average distance from $S$ for each point in dataset, which is an expensive operation (given the number of points are close to $2.5$ million). As an alternative to this, a candidate set is queried first by combining the $k$-nearest neighbours from each of the initial adopters in $S$. Then the candidates are ranked based on their average distance from $S$. Thus, this procedure amounts to finding the most similar users of the current adopters in $S$ and ranking them based on their average distance from $S$ to extract top-$k$ users.
\end{itemize}

\subsubsection{Experiments}
The total number of topics (hashtags) in dataset are 3,617,312. For training user representations, 80\% of topics and their adoption sequences (time-ordered list of users tweeting the hashtag) were used and the rest 20\% topics were held-out for testing. User vectors for 2,574,807 users were obtained from the train set. Vectors were normalised to have unit length, after training. For evaluation, we considered 100 randomly sampled topics from the held-out set which have at least 500 adopters. In case if a user adopts (tweets) a topic (hashtag) multiple times then only the first adoption is taken.\\
For ranking candidates, we used the average based score as it performed better than the other in the experiments. 
Two baseline methods were considered for comparison,
\begin{description}
\item[Frequency Rank] returns users in decreasing order of their frequency of adoption as observed in training topics. Thus, it predicts the same users irrespective of the topic.
\item[Exposure Rank] ranks users according to the number of possible exposures to the topic from their neighbours, which is given by number of following links from initial adopters. This is based on the empirical observation that the likelihood of adoption of a user increases as the number of adopters in its neighbourhood increases (\cite{bakshy2009social} ,\cite{romero2011differences}).
\end{description}
For evaluation the metric used is $Precision@k$, for $k=10$, i.e., the fraction of candidates in the top-\textit{k} list which have been correctly predicted as adopters. Average of the $Precision@10$ values for the topics in test set are reported.

\subsubsection{Results}
\begin{figure}[ht!]
\centering 
\includegraphics[width=0.65\columnwidth]{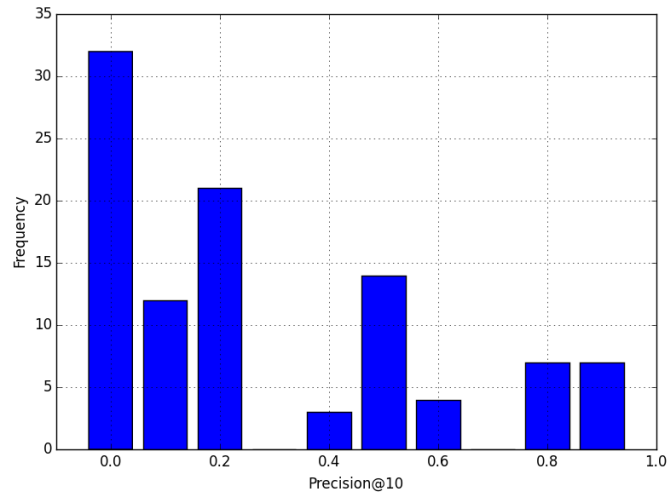} 
\caption[Histogram of Precision@10 values]{Histogram of Precision@10 values for 100 topics, $n=10$}
\label{fig:histprec} 
\end{figure}

We compare the prediction performance of the proposed approach with the baselines in Table~\ref{tab:predadop}. It is observed that with increase in the number of initial adopters, the accuracy of prediction also increases in most of the cases. Moreover, the proposed method outperforms the baselines. Also, there is an increase in performance when vectors of dimension 300 are used, indicating that the added dimensions increases the amount of information captured by the user vectors. Figure~\ref{fig:histprec} plots histogram of Precision@10 values for $n=10$. There is a large variation in predictive performance among topics.

\begin{table}[ht!]
\centering
\tbl{Results of adopter prediction task, $n$ is number of initial adopters\label{tab:predadop}}{
\begin{tabular}{|l|l|l|l|l|l|}
\hline
\begin{tabular}[c]{@{}l@{}}{\bf Model/ No. of initial adopters}\\ {\bf (Precision@10)}\end{tabular} & {\bf n=10}  & {\bf n=20}  & {\bf n=30}  & {\bf n=40}  & {\bf n=50}  \\ \hline
Frequency rank                                                                         & 0.089 & 0.089 & 0.088 & 0.087 & 0.086 \\ \hline
Exposure rank                                                                          & 0.174 & 0.195 & 0.19  & 0.20  & 0.193 \\ \hline
\begin{tabular}[c]{@{}l@{}}User vectors\\ (100-dimensional vector)\end{tabular}        & 0.303 & 0.304 & 0.317 & 0.312  & 0.32 \\ \hline
\begin{tabular}[c]{@{}l@{}}User vectors\\ (300-dimensional vector)\end{tabular}        & 0.352 & 0.377 & 0.394 & 0.386 & 0.398  \\ \hline
\end{tabular}}
\end{table}

\section{Conclusion}
\label{sec:conclusion}

In this paper we have attempted to build user representations based on
their activity on Twitter. Specifically we took collections of tweets
which all contained the same hashtag and used time-ordered sequences
of the tweets in each collection to provide contexts in the sense that
the set of users who tweeted on a particular hashtag proximately in
time to a given user were considered the context of that user. We used
the {\em word2vec} technology that has been used in the context of text
mining to train our model and compute representations of users. 

We were able to use these representations to derive some interesting
insights. We learned that the network neighborhood of a user (in the
sense of followers and followings) is not as similar to the user as
the users more proximate in the representation when it comes to their
activity patterns and their preference for certain hashtags. This
shows that an activity-based view of Twitter yields a different view
than a topology-based view. 

The power of the representations we computed was further demonstrated
by the fact that we were able to use them to predict the geolocation
of a given user with accuracy superior to the baselines considered. It
is notable that no geolocation information was used to construct the
representations. What was more in line with our expectations was the
good performance of the representations on the task of topic adopter
prediction. 

It is our belief that user representations are an important and
powerful tool for extracting information from activity data in complex
settings like online social networks. This paper is an attempt to
begin fleshing out the methodological issues associated with this line
of research and demonstrating the promise this technique holds.

\begin{acks}
The authors would like to thank Siddharth Bora for providing the dataset used in this work.
\end{acks}

\bibliographystyle{ACM-Reference-Format-Journals}
\bibliography{main-bibfile}

\received{February 2007}{March 2009}{June 2009}

\end{document}